\documentstyle[prd,twocolumn,eqsecnum,aps]{revtex}

\begin{document}

\draft
\twocolumn[\hsize\textwidth\columnwidth\hsize\csname
@twocolumnfalse\endcsname

\title{Uniqueness of Self-Similar Asymptotically
Friedmann-Robertson-Walker Spacetime in Brans-Dicke theory}
\author{Hideki~Maeda$^1$
\footnote{Electronic address: hideki@gravity.phys.waseda.ac.jp},
  Jun-ichirou~Koga$^2$
\footnote{Electronic address: koga@gravity.phys.waseda.ac.jp},
and Kei-ichi~Maeda$^{1,2,3}$
\footnote{Electronic address: maeda@gravity.phys.waseda.ac.jp}\\~}
\address{$^1$Department of Physics, Waseda University, Okubo 3-4-1,
Shinjuku, Tokyo 169-8555, Japan\\[-1em]~}
\address{$^2$ Advanced Research Institute for Science and Engineering,
Waseda University, Shinjuku, Tokyo 169-8555, Japan\\[-1em]~}
\address{$^3$ Waseda Institute for Astrophysics,
Waseda University, Shinjuku, Tokyo 169-8555, Japan~}

\date{\today}
\maketitle

\begin{abstract}
We investigate spherically symmetric self-similar solutions in
Brans-Dicke theory. Assuming a perfect  fluid with the equation of state
$p=(\gamma-1)\mu~(1 \le \gamma<2)$, we show that there are no non-trivial 
solutions
which approach asymptotically to the flat Friedmann-Robertson-Walker 
spacetime if
the energy density is positive. This result suggests that primordial black 
holes in
Brans-Dicke theory cannot grow at the same rate as the size of the 
cosmological particle horizon.
\end{abstract}

\vskip 1pc
]

\section{Introduction} It has been pointed out that black holes could be 
formed in
the early universe as a result of the collapse of the initial density
perturbations, bubble collisions, or other
mechanisms~\cite{zn1967,hawkingpbh,pbhform}. They are called primordial 
black holes
(PBHs) and are of special interest since they could be the promissing 
candidates of
MACHO or the only ones which could be small enough to be evaporating by 
the  present
epoch due to the quantum effects~\cite{hawkingrad}. The evolution of PBHs 
has been
studied by many authors in general
relativity~\cite{ch1974,lcf1976,bh1978a,bh1978b}.

Zeldovich and Novikov pointed out the catastrophic growth of the PBHs in a 
simple
Newtonian argument which implies that the PBHs of which size is comparable 
to the
cosmological particle horizon could continue to grow at the same rate as it
throughout the radiation era, while those of which size is much smaller 
than that
could not grow much at all~\cite{zn1967}. One may think that the evolution 
of such
PBHs could be described by self-similar solutions in general relativity. 
Carr and
Hawking, however, showed that there is no non-trivial spherically symmetric
self-similar solution approaching asymptotically to the flat
Friedmann-Robertson-Walker (FRW) spacetime at large radius for a radiation
fluid~\cite{ch1974}. This argument was extended to the case of a general 
fluid with
$p=(\gamma-1)\mu~(1<\gamma<2)$~\cite{bh1978b}. The result implies  that
the PBHs must soon become much smaller than the size of the cosmological
particle  horizon
and cannot grow very much at all.

It has also been argued that in the early universe, gravity might obey a
scalar-tensor type theory rather than general relativity. A scalar-tensor 
type theory of gravity arises naturally as a low-energy limit of string
theory.  In such
theories, gravitational ``constant'' is given by a function of a scalar 
field which
couples non-minimally with gravity so that the effective gravitational 
``constant''
$G$ generally varies in space or in time. Brans-Dicke theory is the 
simplest but
still very important one among the scalar-tensor theories. Then, the 
evolution of
PBHs in Brans-Dicke theories has been studied by several
authors~\cite{barrow1992,sst1995,bc1996,jacobson1999,cg2000,ss2000sb2001,hgc2001}.

Barrow proposed two
scenarios for the evolution of  the PBHs in Brans-Dicke 
theory~\cite{barrow1992}.
In his first scenario (A), the Brans-Dicke scalar field evolves everywhere
homogeneously so that the gravitational ``constant'' at the black hole event
horizon always the same as that at the cosmological particle horizon. No
information about formation epoch remains. In his second scenario (B), however,
the
gravitational ``constant'' at the black hole event horizon remains constant 
with
time and is therefore always the same as that at the formation epoch, while it
changes at distant scales. Since the gravitational ``constant'' at the horizon
stays an old value when a black hole was formed, the remaining gravitational
``constant'' near a PBH is called the {\it gravitational memory}. 
Discussing the
energy emission due to the Hawking radiation in these scenarios, Barrow and 
Carr
found that there are the significant deviations from that in general 
relativity in
both scenarios~\cite{bc1996}. Two alternative scenarios, which are intermediate
between the scenario (A) and (B), are proposed by Carr and Goymer\cite{cg2000}.
  Analyzing perturbations of a scalar field in the
Schwarzschild background~\cite{jacobson1999}, Jacobson  concluded that the
gravitational memory is weak. However, this
conclusion does not follow when the size of a black hole is comparable to 
that of
the cosmological particle horizon.  Using the Tolman-Bondi solution and 
neglecting
  back reaction of a scalar field to the background geometry, Harada,
Goymer and Carr  studied the evolution of a scalar field
when a black hole forms from the collapse of dust in the flat FRW spacetime and
concluded that there is little gravitational memory~\cite{hgc2001}.

The purpose of this short note is to study the evolution of the PBHs in Brans-Dicke
theory by using self-similar solutions which contain a perfect fluid with the
equation of state $p=(\gamma-1)\mu~(1 \le \gamma<2)$.  In this paper, we 
adopt the
unit of $c=1$.

\section{Spherically symmetric self-similar solution in Brans-Dicke theory} The
field equations in Brans-Dicke theory are given by

\begin{eqnarray}
  R_{\mu\nu}&-&\frac12g_{\mu\nu}R =
\frac{8\pi}{\phi}T_{\mu\nu}
+\frac{1}{\phi}(\phi_{;\mu;\nu}-g_{\mu\nu}\Box
\phi)
\nonumber \\
& &~~~~~~~~~~~+\frac{\omega}{\phi^2}
\left(\phi_{;\mu}\phi_{;\nu}-\frac12g_{\mu\nu}\phi_{;\rho}\phi^{;\rho}\right)
,\label{efe} \\
&&T^{\mu\nu}_{\quad;\nu}=0,\\
&&\Box \phi = \frac{8\pi}{3+2\omega}T^{\mu}_{\mu},
\label{bdeom}
\end{eqnarray}
where the constant $\omega$ is the Brans-Dicke parameter and $\phi$
is the Brans-Dicke scalar field, which is related to the gravitational 
``constant''
by $G=1/\phi$. $\omega>-3/2$ is required for the theory not to contain 
tachyonic
solutions.

  We consider a
perfect fluid as a matter field
$ T_{\mu\nu} = p g_{\mu\nu}+(\mu+p) U_{\mu} U_{\nu},
$ where $U_{\mu}$ is the 4-velocity. The radial coordinate $r$ is chosen to be
comoving. The equation of state is assumed to be
$p = (\gamma-1)\mu$ where $1 \le \gamma<2$.
A solution of the Einstein equations is said to be self-similar if it admits a
homothetic Killing vector
${\bf\xi}$ such that
${\cal{L}}_{\bf\xi} g_{\mu\nu} = 2g_{\mu\nu}$, where ${\cal{L}}_{\bf\xi}$ 
denotes
Lie derivative along ${\bf\xi}$.
Cahill and
Taub~\cite{ct1971} first investigated spherically symmetric self-similar 
solutions
in which the homothetic Killing vector is neither parallel nor orthogonal 
to the
fluid flow vector. They showed that by the suitable coordinate 
transformations such
solutions can be put in a form in which all dimensionless quantities are 
functions
only of the dimensionless variable
$z \equiv r/t$.
Generalizing their method to the present model, we find the following basic
equations.
The line element of a spherically symmetric self-similar spacetime is given by
\begin{eqnarray}
ds^2 &=& -e^{2\Phi(z)}dt^2+e^{2\Psi(z)}dr^2+ r^2 S^2(z) ~d\Omega^2,
\end{eqnarray}
where $d\Omega^2=d\theta^2+\sin^2 \theta d\varphi^2$.
Here we assume the self-similarity of a perfect fluid and the
Brans-Dicke scalar field independently, i.e.

\begin{eqnarray}
{\cal{L}}_{\bf\xi}p = a p,\quad  {\cal{L}}_{\bf\xi}\mu = b
\mu,\quad {\cal{L}}_{\bf\xi}\phi = \kappa
\phi,\label{ssphi}
\end{eqnarray}
where $a,b$ and $\kappa$ are constants. These constants satisfies
the relations $a=b=-(2-\kappa)$ through the equation of state  and
the field equations (\ref{efe})-(\ref{bdeom}). Equation (\ref{ssphi}) implies
that the quantities $p,\mu$ and $\phi$ must be of the form as

\begin{eqnarray}
\mu=\frac{W(z)}{r^{2-\kappa}}, \quad p=\frac{P(z)}{r^{2-\kappa}}, \quad
\phi=r^{\kappa}\psi(z),
\end{eqnarray}
and then equations (\ref{efe})-(\ref{bdeom}) reduce to the
ordinary differential equations w.r.t. the self-similar variable $z$.

The energy-momentum conservation equation
  is written in self-similar
variable as

\begin{eqnarray}
\Phi' &=& \frac{(2-\kappa)P-P'}{W+P} = {\gamma-1\over \gamma}\left(
2-\kappa +{W'\over W}\right), 
\label{ece1}\\
 \Psi' &=&
-\left(\frac{W'}{W+P}+2\frac{S'}{S}\right)
= -\left({1\over \gamma}{W'\over  W}+2\frac{S'}{S}\right), \label{ece2}
\end{eqnarray} where prime denotes the derivative with respect to $\ln z$. 
These
equations can be integrated to give

\begin{eqnarray} e^{\Phi}&=&c_0
z^{\frac{(2-\kappa)(\gamma-1)}{\gamma}}W^{-\frac{\gamma-1}{\gamma}},
\label{integral1}\\ e^{\Psi}&=&c_1 
S^{-2}W^{-\frac{1}{\gamma}},\label{integral2}
\end{eqnarray} where $c_0, c_1$ are integration constants. Equations
(\ref{efe}) and (\ref{bdeom}) then reduce to the following ordinary 
differential
equations;


\begin{eqnarray}
&&\left[{(2+\kappa)(\gamma-1)\over \gamma}\left((2-\kappa)
-\frac{W'}{W}\right)-(\kappa-1
+\omega\kappa)\right.
\nonumber
\\
&&\left.
\times \left(\kappa+\frac{\psi'}{\psi}\right)\right]
-V^2\left[-(2+\kappa)\left(2\frac{S'}{S}
+\frac{1}{\gamma}\frac{W'}{W}\right)
\right.
\nonumber
\\
&&
\left.
-(\kappa-1
+\omega\kappa)\frac{\psi'}{\psi}\right]
=\frac{8\pi\gamma }{\psi}c_1^2S^{-4}W^{1-\frac{2}{\gamma}},\label{00+11}
\\[1em]
&&
2\left(\frac{S''}{S}+\frac{S'}{S}\right)
+\frac{\psi''}{\psi}+\kappa\frac{\psi'}{\psi}
+\omega\frac{\psi'}{\psi}\left(\kappa+\frac{\psi'}{\psi}\right)
+2\frac{S'}{S}\frac{\psi'}{\psi}
\nonumber
\\
&&
-\left({(2-\kappa)(\gamma-1)\over \gamma}
-\frac{W'}{W}\right)\left(2\frac{S'}{S}+\frac{\psi'}{\psi}\right)
\nonumber
\\
& &
+\left(2\frac{S'}{S}
+\frac{1}{\gamma}\frac{W'}{W}\right)
\left[\kappa+2\left(1+\frac{S'}{S}\right)\right]=0,\label{01}
\\[1em]
& &
\frac{\psi''}{\psi}+(2\kappa-1)\frac{\psi'}{\psi}+\kappa(\kappa-1)
\nonumber
\\
& &
+\left(\kappa+\frac{\psi'}{\psi}\right)\left(2
+\frac{(2-\kappa)(\gamma-1)}{\gamma}+4\frac{S'}{S}
-\frac{\gamma-2}{\gamma}\frac{W'}{W}\right)
\nonumber
\\
& & -V^2 \left[\frac{\psi''}{\psi}+\frac{\psi'}{\psi}\left(1
-\frac{(2-\kappa)(\gamma-1)}{\gamma}
+\frac{\gamma-2}{\gamma}\frac{W'}{W}\right)\right]
\nonumber
\\
& &=
\frac{8\pi(3\gamma-4)}{(3+2\omega)\psi}~c_1^2S^{-4}W^{1-\frac{2}{\gamma}},
\label{eomphi}
\end{eqnarray}
with $V$ $\equiv$ $z
e^{\Psi-\Phi}$=$(c_1/c_0)z^{1-[(2-\kappa)(\gamma-1)/
\gamma]}$ $S^{-2}$ $W^{(\gamma-2)/\gamma}$. Equations (\ref{00+11}) and (\ref{01}) are independent of equations (\ref{ece1}) and (\ref{ece2})

\section{The flat FRW solution}

The flat FRW solution in Brans-Dicke theory with a perfect fluid obeying the
equation of state $p=(\gamma-1)\mu$ was found by Nariai~\cite{nariai1968} 
and it is
a particular solution of the reduced ordinary differential equations, i.e. a
self-similar solution. The flat FRW solution in the self-similar variable
($\Phi_{FRW}(z), \Psi_{FRW}(z), S_{FRW}(z), W_{FRW}(z), $)
is

\begin{eqnarray}
e^{\Phi_{FRW}}&=&a_0,\quad e^{\Psi_{FRW}}=b_0z^{-q},\quad
\label{FRWsolution1}\\
S_{FRW}&=&\frac{b_0}{1-q}z^{-q},\quad
\psi_{FRW}=\psi_0 z^{-2+3q\gamma}, \\
W_{FRW}&=&
{\psi_0(3+2\omega)[2(5-3\gamma)+3(2-\gamma)^2\omega] \over
4\pi{a_0}^2
[4-3\gamma\omega(\gamma-2)]^2]}
~z^{3q\gamma},
\label{FRWsolution3}\\
\kappa &\equiv& \frac{2(4-3\gamma)}{4-3\gamma\omega(\gamma-2)},
~~~~
q \equiv
\frac{2[1-\omega(\gamma-2)]}{4-3\gamma\omega(\gamma-2)}. \label{q}
\end{eqnarray}
One can put this solution in a more familiar form 
\begin{eqnarray}
ds^2=-dt'^2+t'^{2q}(dr'^2+r'^2 d\Omega^2),
\end{eqnarray}
by making the coordinate
transformation
$t'=a_0t,r'=a_0^{-q}b_0r^{1-q}/(1-q)$.
In the case of radiation ($\gamma=4/3$),
$q=1/2$ and $\kappa=0$ are obtained from equation (\ref{q}). Since $\kappa=2-3q\gamma$, $\kappa=0$ implies that
the Brans-Dicke  scalar
field $\phi$ is constant which case is equivalent to that in general 
relativity.
  Assuming the energy density is positive ($W>0$), we find
\begin{eqnarray}
\omega >\frac{2(3\gamma-5)}{3(\gamma-2)^2} = -{3\over 2} +{(3\gamma-4)^2\over
6(\gamma-2)^2}.
\end{eqnarray}
If $\gamma=2$, any value of $\omega$ is not possible.
On the other hand, in the case of radiation ($\gamma=4/3$), the energy density
is always
positive for any value of
$\omega>-3/2$.

\section{Uniqueness of self-similar asymptotically flat FRW solution}

We are interested in self-similar solutions which are asymptotically flat FRW
solution, i.e. in which $W, S$ and $\psi$ approach the form
(\ref{FRWsolution1})-(\ref{FRWsolution3}) as $z \to \infty$. It is 
convenient to
introduce new functions $A,B$ and $C$, which denote deviations from the 
flat FRW
solution, defined by

\begin{eqnarray}
W&=&W_{FRW}(z) e^{A(z)},\label{defA}
\quad  \\
S&=&S_{FRW}(z) e^{B(z)}, \quad \psi=\psi_{FRW}(z)e^{C(z)}.\label{defBC}
\end{eqnarray}
These functions denote deviations from the flat FRW solution, and equations (\ref{defA}) and (\ref{defBC}) imply (through equations (\ref{ece1}) and (\ref{ece2})) that

\begin{eqnarray}
e^{\Phi}&=&e^{\Phi_{FRW}}e^{-\frac{\gamma-1}{\gamma}A},\quad
e^{\Psi}=e^{\Psi_{FRW}}e^{-2B-\frac{A}{\gamma}}.
\end{eqnarray}

  The flat FRW solution is given by $A=B=C=0$ for any $z$.
Since $
V=(b_0/a_0)z^{1-q}\exp(-2B+\frac{\gamma-2}{\gamma}A)
$, the
asymptotic form of self-similar solutions which approach the FRW solution for
large
$z$ can be found by linearizing equations  neglecting the $V^2$
term for accelerating expansion case ($q>1$), while the $1/V^2$ term for
deaccelerating expansion case ($0<q<1$).

\subsection{Accelerating expansion case}

For accelerating case ($q>1$), the asymptotic behavior of linearized 
equations are
given by
\begin{eqnarray}
C'' &=& -(1-q)C',\\
  2B'' &=&-C''+\frac{1}{\gamma}(\kappa\gamma+2q\gamma-2-\kappa)A' \nonumber \\
&&+6(q-1)B'+(-q+\kappa+\kappa\omega)C',\\
  A'
&=& -{\gamma(\kappa-1+\omega\kappa)\over (\gamma-1)(2+\kappa)}C'.
\end{eqnarray}
This system of linear differential equations has no solution which vanish as $z \to \infty$. It means that there are no non-trivial asymptotically flat FRW solutions in accelerating case.

\subsection{Deaccelerating expansion case}
For deaccelerating case ($0<q<1$)
the asymptotic behavior of linearized equations are
given by
\begin{eqnarray}
-C'' &=&
-\frac{3\gamma-4}{3+2\omega}\left(3q^2-\frac{\omega}{2}\kappa^2
+3q\kappa\right)\left(C+\frac{\gamma-2}{\gamma}A\right)
\nonumber \\ &&+(-2\kappa+1-3q)C'-\frac{\kappa(\gamma-2)}{\gamma}A',\\ 2B'' &=&
-C''+\frac{1}{\gamma}(\kappa\gamma+2q\gamma-2-\kappa)A'
\nonumber \\ &&+6(q-1)B'+(-q+\kappa+\kappa\omega)C',\\
\frac{2+\kappa}{\gamma}A' &=&
-\gamma\left(3q^2-\frac{\omega}{2}\kappa^2+3q\kappa\right)\left(C
+\frac{\gamma-2}{\gamma}A\right)
\nonumber \\ &&-2(2+\kappa)B'-(\kappa-1+\omega\kappa)C',
\end{eqnarray}
which solutions is
\begin{eqnarray}
C&=&\frac{2-\gamma}{\gamma}A+\frac{C_0}{z^s}, \quad
3B+\frac{A}{\gamma}=\frac{C_1}{z^s}, \quad A=\frac{C_2}{z^s}, \\
&&s \equiv
1-3q-\kappa=
-{2-\gamma\over \gamma}{\left[\omega-{2(3\gamma-5)\over 
3(\gamma-2)^2}\right]\over
\left[\omega  +{4\over 3\gamma(2-\gamma)}\right]},
\end{eqnarray}
where $C_0,C_1$ and $C_2$ are constants satisfying one constraint equation
\begin{eqnarray}
&&-\left[\omega(3\gamma^2-8\gamma+8)+4\right]C_0
+\frac43\left[3\omega\gamma(\gamma-2)+3\gamma-8\right]C_1
\nonumber \\ &&
+\frac{3\gamma-4}{3\gamma}\left[3\omega(\gamma-2)^2-2(3\gamma-5)\right]C_2=0.
\end{eqnarray}
The power $s$ is positive when
\begin{eqnarray}
\left(\omega+\frac{4}{3\gamma(2-\gamma)}\right)
\left(\omega-\frac{2(3\gamma-5)}{3(\gamma-2)^2}\right) <0.
\label{positive_s}
\end{eqnarray}
  When $\gamma=4/3$, $s=-1/2<0$. Since for deaccerelating expansion ($0<q<1$),
we find
\begin{eqnarray}
\omega &>&-{2\over (2-\gamma)(3\gamma-2)}>-\frac{4}{3\gamma(2-\gamma)} 
~~~{\rm for
}~~~\gamma>{4\over 3},
\nonumber \\
\omega &>&-{1\over 2-\gamma}>-\frac{4}{3\gamma(2-\gamma)}  ~~~{\rm
for }~~~\gamma<{4\over 3}, 
\label{deacceralation}
\end{eqnarray}
the inequality (\ref{positive_s}) with (\ref{deacceralation}) yields that
  the energy density of a perfect fluid is negative.
Then the positivity of energy density requires $s<0$, i.e., there is no non-trivial solution which approaches
the flat FRW universe.
We conclude that there are no non-trivial asymptotically flat
FRW solutions in deaccelerating case.

\section{Discussion and Summary} We have studied spherically symmetric 
self-similar
solutions in Brans-Dicke theory which contain a perfect fluid with the 
equation of
state $p=(\gamma-1)\mu~(1 \le \gamma<2)$. We have shown that when the energy
density is positive, there are no non-trivial solutions which approach to 
the flat
FRW spacetime at large $z$. It should be noted that although non-trivial solutions do exist,
they are unphysical solutions with negative energy density. It is similar to the result in general relativity. In general relativity, non-trivial solutions which approach to the flat FRW spacetime at large $z$ exist for $\gamma>2$, where the sound speed is faster than the speed of light~\cite{ch1974}.

The result in this paper suggests that the PBHs does not 
grow at the
same rate as the size of the cosmological particle horizon in Brans-Dicke 
theory as
in the case of general relativity. However, it could be possible to construct
a non-trivial solution attached to the flat FRW solution at the point of
$V^2=1$ or $V^2=\gamma-1$, where the field equations become singular.
  In general
relativity, it was shown that the ingoing Vaidya null fluid solution can be
attached to the flat FRW solution at the point of $V^2=\gamma-1$ if the 
equation
of state is $p=\mu$ though this is not a black hole solution~\cite{bh1978a}.

If we allow deficit (surplus) angle in the present self-similar solutions, 
we can
find  a one-parameter family of solutions which approaches asymptotically 
to the
``flat FRW spacetime", as
  Carr and Hawking found them
in the case of general relativity~\cite{ch1974}.

In this paper, we have imposed the self-similarity on the whole spacetime, 
however
it can be considered that the self-similarity is imposed only on the 
neighborhood
of a black hole and the solution smoothly connects to the non-self-similar
spacetime at large distance. Such a solution represents a black hole in the flat FRW universe which grows locally in the self-similar manner. To construct such a solution is an important future work.

\acknowledgements

We are very grateful to N.~Inada for discussions and comments. This work was  
partially by the Grant-in-Aid for Scientific Research Fund of the Ministry of
Education, Science and Culture (Specially Promoted Research No. 14047216) and by the
Waseda University Grant for Special Research Projects.

\end{document}